\documentclass[prl,priprint,epsf,psfig]{revtex4}
\usepackage{graphicx}
\DeclareGraphicsExtensions{.pdf,.png,.gif,.jpg}
\usepackage{float}
\usepackage{subfig}
\usepackage{epsfig}
\usepackage{wrapfig}

\begin{document}
\title{An effective mean field theory for the coexistence of anti-ferromagnetism and superconductivity: Applications to iron-based superconductors and cold Bose-Fermi atomic mixtures}

\author{Jeremy Brackett}
\thanks{These two authors contributed equally.}
\author{Joseph Newman}
\thanks{These two authors contributed equally.}
\author{Theja N. De Silva}
\affiliation{Department of Chemistry and Physics, Augusta University, Augusta, GA 30912, USA.}
\begin{abstract}

We study an effective fermion model on a square lattice to investigate the cooperation and competition of superconductivity and anti-ferromagnetism. In addition to particle tunneling and on-site interaction, a bosonic excitation mediated attractive interaction is also included in the model. We assume that the attractive interaction is mediated by spin fluctuations and excitations of Bose-Einstein condensation (BEC) in electronic systems and Bose-Fermi mixtures on optical lattices, respectively. Using an effective mean-field theory to treat both superconductivity and anti-ferromagnetism at equal footing, we study a single effective model relevant for both systems within the Landau energy functional approach and a linearized theory. Within our approaches, we find possible co-existence of superconductivity and anti-ferromagnetism for both electronic and cold-atomic models. Our linearized theory shows while spin fluctuations favor d-wave superconductivity and BEC excitations favor s-wave superconductivity.
\end{abstract}

\maketitle

\section{I. Introduction}

The phenomenon of superconductivity has been an important and rich topic in physics since 1911 when Kamerlingh Onnes discovered that the resistivity of mercury abruptly dropped to zero when it was cooled below 4K~\cite{onnes}. Over the years, a number of superconducting compounds were found or grown with higher critical temperatures. The highest critical temperature achieved in this class of conventional superconductors was 40 K in magnesium diboride~\cite{budko}. If one can maintain this resistanceless superconducting state at room temperature, society would obtain huge economic benefits as these compounds can be used to store and transport energy without dissipation. In metal, even though electrons are free to move and provide electrical conduction, energy dissipation occurs due to the resistance coming from electron collisions, lattice vibrations, impurities, and defects. The resistanceless state of these conventional superconductors is explained by the celebrated BCS theory developed by Bardeen, Cooper, and Schrieffer in 1957~\cite{bcs}. According to the BCS theory, two electrons with equal and opposite speed bind together due to the attractive interaction mediated by the electron-phonon interaction. These bound pairs are called Cooper pairs. As the Cooper pairs are composite bosons, made out of two fermions, Bose-Einstein condensation of these pairs at low temperatures gives resistanceless flow.

The discovery of cuprate superconductors in 1986 by Bednorz and Muller~\cite{cuprates1} has renovated the interest of superconductivity as these compounds have quite high critical temperatures so they may be useful in practical applications. Immediately after this discovery, several other cuprate superconducting compounds with higher critical temperatures were found~\cite{cuprates2, cuprates3, cuprates4, HTCCO}. Cuprates are considered to be quasi-two dimensional checkerboard lattice materials as the electrons are moving within weakly coupled copper-oxide layers. The highest critical temperature of cuprates at ambient pressure so far is 135 K in mercury barium calcium copper oxide~\cite{HTCCO}. Then the surprising discovery of iron-based superconductors in 2008 has led to a flurry of activities in the field as these compounds provide more puzzles than answers to the questions of unconventional superconductivity~\cite{pnictide}. The critical temperatures of iron-based superconductors are in between that of conventional superconductors and cuprates. The iron based superconductors share some common features with the cuprates. Both are layered materials with $3d$-electrons. Iron-based superconductors contain layers of iron and pnictogen (arsenic or phosphorus) or chalcogen. Both cuprates and iron based superconductors require chemical or external doping to induce the superconductivity.  One of the main differences between these two types of compounds is that the orbital degrees of freedom associate with the Fe-ion in iron-based superconductors. Iron based superconductors are essentially multi-orbital systems so that the electron occupation of the $d$-orbital must be taken into account. In contrast, cuprates can be treated as single orbital systems as the crystal field splitting of $d$-orbital and valence electronic occupation restrict one hole in the upper most $d$-orbital.

For both cuprates and iron-pnictides, critical temperatures are too high to be explained by conventional BCS theory. Therefore, the effective interaction for electron pairing must be mediated by excitations rather than conventional phonons. Though it is not completely convincing, there is a general consensus that magnetic or spin fluctuations play the role as the pairing mechanism for these compounds~\cite{sp}. However, the role of magnetism, the nature of chemical and structural influence, and the pairing symmetry of the electrons are not completely understood. It has been shown that the s-wave pairing is suppressed by ferromagnetic spin fluctuations~\cite{sfmfluc}. However, p-wave pairing  can be enhanced by the ferromagnetic spin fluctuations~\cite{ pfmfluc}. In both cuprate and pnictide superconductors, the superconductivity always appears in proximity to anti-ferromagnetic (AFM) order or they co-exist in some compounds upon chemical doping. This indicates that the Cooper pairing may be mediated by anti-ferromagnetic fluctuations.

The purpose of this paper is to study the interplay between induced interactions and superconductivity, as well as the competition or cooperation of anti-ferromagnetism and superconductivity. For this purpose, we study both high-temperature superconducting materials and cold atoms in optical lattices. Cold atoms on optical lattices can be considered as quantum simulators for condensed matter electronic systems. One important advantage of using optical lattices to probe fundamental condensed-matter physics problems is that the geometry, dimensionality, and the interaction parameters are under complete control in current experimental setups~\cite{CAReview1, CAReview2, CAReview3}. This high degree of tunability and controllability offers a remarkable opportunity to understand and fully explore the quantum mechanical treatment of cold atomic systems whose behavior is governed by the same underlying many body physics as the materials. In this paper, in addition to the layered high-temperature compounds, we consider a mixture of bosons and two-component fermions in a two-dimensional optical lattice. In experiments, atoms are trapped by combined harmonic trapping and periodic laser potentials. The periodic potentials are created by the interference patterns of intersecting laser beams, and the geometry of the lattice structure can be controlled by the arrangements of the counter propagating lasers. Bose-Fermi mixtures have already been trapped and experimentally studied by several groups~\cite{bfm1, bfm2, bfm3, bfm4, bfm5, bfm6, bfm7, bfm8, bfm9, bfm11, bfm12, bfm13}. A Bose-Fermi mixture, such as a mixture of $^{41}$K and two hyperfine states of $^6$Li or a mixture of $^6$Li$^{40}$K bosonic molecule and fermionic species $^{40}$K and $^6$Li is an example of two-component Fermions and single-component bosons system we study here. Such systems have already been experimentally realized in different settings~\cite{bfop1, bfop2, bfop3, bfop4}. For high-temperature materials, such as cuprates and pnictides, we assume that the effective interaction is caused by spin fluctuations~\cite{spf1, spf2}. For the Bose-Fermi mixture, we assume that the effective attractive interaction is mediated by the density fluctuations of the bosonic atoms~\cite{bdf1, bdf2, bdf3}. In both cases, the Cooper pairing between Fermi particles can take place due to these boson mediated attractive interactions.

The interplay between superconductivity and anti-ferromagnetism has been investigated in the context of cuprates, organic superconductors, heavy fermion systems~\cite{scafC1, scafC2, scafC3, scafC4, scafC5, scafC6}, and iron based superconductors~\cite{scafC7, scafC8, scafC9, scafC10, scafC11, scafC13, scafC14}. Using a combination of renormalization group and mean-field theory, the competition and coexistence of d-wave superconductivity and antiferromagnetism in the ground state of the two-dimensional Hubbard model has been studied in the context of the cuprates recently~\cite{nt1a, nt1b}. The results of this study is in good agreement with the early findings from the dynamical mean field theoretical studies of the ground state of Hubbard model~\cite{nt2, nt3, nt4, nt5}. In this work, we develop a simple mean field theory to understand the qualitative physics of the finite temperature coexistence of super conductivity and anti-ferromagnetism relevant for a electronic system and a Bose-Fermi atomic mixture on optical lattices.

The study of anti-ferromagnetism and superconductivity in a lattice model discussed here is somewhat complementary to early studies~\cite{scafC2, oldt1, oldt2} and recent studies related to cold atoms~\cite{oldt3, oldt4, oldt5, oldt6, oldt7}. However, unlike those studies which assume a generic form of interaction, here we treat explicit momentum dependent interaction relevant for both Bose-Fermi mixtures on optical lattices and related electronic model for iron-based superconductors. Further, we treat both s-wave and d-wave superconducting symmetries at equal footing with their interplay between anti-ferromagnetism. In addition, we point out how one can control the anti-ferromagnetic phase transition to be below or above the superconducting phase transition by controlling the boson density in Bose-Fermi mixtures. In order to study the interplay between superconductivity and anti-ferromagnetism, we develop an effective mean-field theory for an effective fermionic Hamiltonian relevant for both electronic compounds and Bose-Fermi mixtures on two-dimensional lattices. First, we investigate the phase diagram using the Landau energy functional and derive coefficients of this energy functional within our mean-field theory. Second, we study the phase transition by solving the linearized gap equations. For both electronic and atomic systems, we find simultaneous existence of superconductivity and anti-ferromagnetism.

The paper is organized as follows. In section II and III, we review the boson mediated attractive interactions for Bose-Fermi mixtures and electronic models, respectively. We devote section IV to discuss our effective mean-field theory for the derivation of thermodynamic potential. In section V, we introduced the Landau energy functional and then in section VI, we derive the gap equation and discuss our linearization scheme. In section VII, we discuss the generic two order parameter phase diagram within the Landau energy functional and derive Landau energy functional coefficients for both electronic and atomic models. In section VIII, we discuss the critical temperatures and phase transitions using our linearized gap equations. Finally in section IX, we summarize the results and provide a general discussion.

 \section{II. Elementary excitation induced attractive interaction between fermions in Bose-Fermi mixture}

In this section, we briefly review the effective interaction between fermions originated from the bosonic density fluctuations. When the lattice potential is strong, the atomic system in the  two-dimensional (2D) square lattice can be modeled by the single band Hubbard model~\cite{jks},

\begin{eqnarray}
H_{bf} = -t_b \sum_{\langle i, j\rangle} \biggr(b_i^\dagger b_j + h. c\biggr) - t_f \sum_{\langle i, j\rangle, \sigma} \biggr(c_{i, \sigma}^\dagger c_{j, \sigma} + h. c\biggr) \\ \nonumber
+ \frac{U_{bb}}{2} \sum_i n_i^b (n_i^b-1) + U_{bf} \sum_i n_i^bn_i^f + U_{ff} \sum_i n_{i, \uparrow} n_{j, \downarrow},
\end{eqnarray}

\noindent where $t_\alpha$ is the boson ($\alpha = b$) and fermions ($\alpha = f$) tunneling amplitudes between neighboring sites $i$ and $j$, respectively. The on-site boson-boson, boson-fermion, and fermion-fermion interactions are denoted by $U_{bb}$, $U_{bf}$, and $U_{ff}$, respectively. The operators $b_i (b_i^\dagger)$ are the on-site bosonic annihilation (creation) operators, and $c_{i, \sigma}$ are the on-site fermionic annihilation operators for pseudo spin $\sigma = \uparrow, \downarrow$. The Bosonic occupation number operator is $n_i^b = b_i^\dagger b_i$ and the fermionic occupation number operator for $\sigma$ spin is $n_{i, \sigma} = c_{i, \sigma}^\dagger a_{i, \sigma}$. The on-site fermionic occupation number operator is then $n_i^f = n_{i, \uparrow} + n_{i, \downarrow}$. The 2D optical lattice is created by three pairs of counter propagating laser beams that provide the trap potential in the form $V_\alpha (\vec{r}) = V_{a\alpha}(z) + V_{t \alpha} (x,y)$ for bosons and fermions with $\alpha = b, f$ respectively, where

\begin{eqnarray}
V_{a \alpha} (z) = V_{\alpha z} \sin^2(\pi z/d) \enspace \hbox{and,} \enspace
V_{t \alpha}(x,y) = V_{\alpha \perp} [\sin^2(\pi x/d) + \sin^2(\pi y/d)].
\end{eqnarray}

\noindent Here, we assume that both bosons and fermions feel the same laser wavelength $\lambda$ which is related to the lattice constant $d$ through $d = \lambda/2$. The quasi-2D structure is maintained by a stronger axial confinement with  $V_{\alpha z} \gg V_{\alpha \perp}$. The tunneling amplitude and the on-site interactions are tunable through transverse lattice strength $V_{\alpha \perp}$ and scattering lengths between two bosons $a_{bb}$, a boson and a fermion $a_{bf}$, and two fermions $a_{ff}$. The tunneling amplitudes are given by~\cite{prms1, prms2}

\begin{eqnarray}
t_\alpha = \frac{4}{\sqrt{\pi}}E_\alpha \biggr(\frac{V_{\alpha \perp}}{E_\alpha}\biggr)^{3/4} \exp\biggr(-2 \sqrt{\frac{V_{\alpha \perp}}{E_\alpha}}\biggr),
\end{eqnarray}

\noindent where $E_\alpha = 2\hbar^2 \pi^2/\lambda^2 m_\alpha$ is the photon recoil energy with mass $m_\alpha$ of a boson or a fermion. The on-site interactions have the form~\cite{prms1, prms2},

\begin{eqnarray}
U_{\alpha \alpha} = \frac{U_{\alpha \alpha}^{3D}}{(2\pi)^{3/2} d_{\alpha z} d^2_{\alpha \perp}} \enspace \hbox{and,} \enspace U_{b f} = \frac{U_{b f}^{3D}}{\pi^{3/2} \sqrt{d^2_{b z}+d^2_{f z}} (d^2_{b \perp}+ d^2_{f \perp})},
\end{eqnarray}

\noindent where $d_{\alpha z, \perp} = \sqrt{\hbar/m_\alpha \omega_{\alpha z, \perp}}$ with $\omega_{\alpha z, \perp} = 2 \sqrt{E_\alpha V_{\alpha z, \perp}}/\hbar$ and three dimensional short-range interactions $U^{3D}_{\alpha \beta} = 4 \pi \hbar^2 a_{\alpha \beta}/m_{\alpha \beta}$ with effective mass $m_{\alpha \beta} = 2 m_\alpha m_\beta/(m_\alpha+ m_\beta)$.

We are interested in the low temperature and strong kinetic energy regime of bosons where the bosonic atoms are condensed in the lattice. When the bosons are in Bose-Einstein condensate, effective interaction between fermions are induced from the elementary excitations of the BEC. As it is well known, these elementary excitations are phonons or sound waves and this phononic excitation spectrum can be calculated from Bogoliubov approximation to the bosons.

In Bogoliubov approximation, one uses the bosonic operators in momentum space as $b_q = \sqrt{N_L n_B} \delta(q) + \tilde{b}_q$ and keeps fluctuation operators $\tilde{b}_q$ up to quadratic order in the bosonic sector of the Hamiltonian. Here $n_B$ and $N_L$ are bosonic density and the number of lattice points, respectively. Then diagonalizing the bosonic sector, one finds the spectrum of elementary excitations of the Bose superfluid~\cite{bgf},

\begin{eqnarray}
\omega_B(k) = \sqrt{\epsilon_B(k) [ \epsilon_B(k) + 2 n_B U_{bb}]}.
\end{eqnarray}

\noindent Here, we have assumed that all the bosons are condensed into the zero momentum state, hence $n_B$ is the superfluid bosonic density. The single particle boson dispersion on the lattice is defined as $\epsilon_B(k) = -2 t_b(\cos k_x d + \cos k_y d)$. Within the same approximation, the boson density-density response in the static limit is then given by,

\begin{eqnarray}
\chi_B(k) = -\frac{2 n_B \epsilon_B(k)}{\omega_B(k)^2}.
\end{eqnarray}

\noindent By integrating out the phonon ($\tilde{b}_q$) field in the effective fermion-phonon coupling Hamiltonian (bosonic part and the boson-fermion coupling term), the phonon mediated attractive interaction between Fermi atoms is given by~\cite{bdf1, bdf2, bdf3},

\begin{eqnarray}
V_{ph}(k) = \chi_B(k) U_{bf}^2.
\end{eqnarray}

\noindent When deriving this effective interaction, renormalization of $\chi_B(k)$ due to the presence of fermions is neglected. This is reasonable as we are considering a dilute Bose-Fermi mixture here. The static limit of the response function can be justified as the resulting interaction between fermions is instantaneous. This is always the case when the velocity of the Bose excitations (phonons) are much larger than the Fermi velocity.

\section{III. Spin fluctuation induced attractive interaction between electrons in high $T_C$ compounds}

In this section, we briefly review the effective interaction between electrons originated from the spin fluctuations. The dynamic of the electron in the  two-dimensional square lattice can be modeled by the single band Hubbard model that includes on-site Coulomb repulsion $U_e$ and nearest neighbor hopping amplitude $t_e$. Here we assume that there is $d$-orbital splitting due to the crystal field effects, Hund's coupling, on-site interaction, and the number of valence electrons in the electronic system are such that the system can be described by the single band Hubbard model. This is certainly the case for cuprates, but various tight-binding models such as two-orbital, three-orbital, and five-orbital have been proposed for pnictides~\cite{models1, models2, models3}. Certainly, the iron based superconducting systems are multi-band systems and multi-band natures, such as Hund's coupling and intra-atomic exchange energies play a role in these materials~\cite{mulo}. However, we believe that the competition between direct on-site interaction and effective attractive interaction relevant for the interplay between anti-ferromagnetism and superconductivity can be studied using a single-band Hubbard model. The model is only a part of the Hamiltonian ($H_{bf}$) presented before, as we have only fermions in the lattice,

\begin{eqnarray}
H_{e} = - t_e \sum_{\langle i, j\rangle, \sigma} \biggr(c_{i, \sigma}^\dagger c_{j, \sigma} + h. c\biggr)  + U_{e} \sum_i n_{i, \uparrow} n_{i, \downarrow} - \mu \sum_{i, \sigma} c_{i, \sigma}^\dagger c_{i, \sigma} .
\end{eqnarray}

\noindent Starting from this Hamiltonian, and adding an external spin-dependent potential (or a magnetic field, which is set to be zero at the end of the calculation) one can use linear response theory to derive the effective interaction between electrons. This spin fluctuation mediated interaction is similar to the elementary excitation mediated interaction discussed in the previous section. In the former case, the bosons responsible for the interaction between electrons are magnetic excitations known as magnons. Even in the paramagnetic state with short-range anti-ferromagnetic order, highly damped magnons mediate interactions between electrons.

Using the exchanges of spin fluctuations within a weak coupling random-phase approximation, the paramagnon mediated effective interaction in the singlet channel is derived using a diagrammatic approach~\cite{sft}

\begin{eqnarray}
V_{mag}(\vec{k}) = \frac{U_e^2 \chi_0(\vec{k})}{1-U_e \chi_0(\vec{k})} + \frac{U_e^3 \chi_0^2(\vec{k})}{1-U_e^2 \chi_0^2(\vec{k})}.
\end{eqnarray}

\noindent Here $\vec{k} = \vec{q}-\vec{q^\prime}$ is the momentum transfer in the scattering of a pair of electrons from state $(\vec{q}, -\vec{q})$ to state $(\vec{q^\prime}, -\vec{q^\prime})$ and $\chi_0(\vec{k})$ is the wave-vector dependent susceptibility of the noninteracting electrons,

\begin{eqnarray}
\chi_0(\vec{q}) = \sum_k\frac{n_f(\epsilon_{k+q}) - n_f(\epsilon_{k})}{\epsilon_k - \epsilon_{k+q}},
\end{eqnarray}

\noindent with the single particle excitation $\epsilon_k = -2t_e(\cos k_x d + \cos k_y d) - \mu$ for electron in the lattice. Here $n_f(x) = 1/[e^{\beta x} +1]$ is the usual Fermi function with dimensionless inverse temperature $\beta = 1/k_BT$. The first term in $V_{mag} (k)$ arises from the transverse spin fluctuations, while the second term corresponds to the longitudinal spin fluctuations~\cite{sft}. Even though the effective interaction $V_{mag} (\vec{k})$ is positive, the $\vec{k}$ dependence plays a major role when it comes to the pairing of electrons. The effective interaction has a peak at $\vec{k} = (\pi/d, \pi/d)$ and its Fourier transform in real space shows an oscillatory behavior between positive and negative values~\cite{ysl}. As a result, two electrons in spatially apart can attract and form a Cooper pair.

\section{IV. An effective mean field theory for the superconductivity and anti-ferromagnetism}

In the low temperature regime where the elementary excitations are dominant, both Bose-Fermi and electronic systems discussed in sections II and III can be represented by an effective fermion Hamiltonian. It is convenient to develop the mean field theory in the momentum representation, \emph{i. e.} we represent the Fermi operators in the plane wave basis as $ c_{i, \sigma} = 1/\sqrt{N_L} \sum_k e^{i\vec{k} \cdot \vec{r}_i} c_{k, \sigma}$, where $N_L$ is the number of lattice sites and $\vec{k}$ runs through the reciprocal lattice. In this Fourier basis, the Hamiltonian for both Bose-Fermi system and electronic system can be written as $H = H_0 + H_{SC} + H_{AF}$,

\begin{eqnarray}
H_0 = \sum_{k, \sigma} \epsilon_k c_{k, \sigma}^\dagger c_{k, \sigma} \\ \nonumber
H_{SC} = \frac{1}{2V_l} \sum_{k k^\prime} V_{k k^\prime} c^\dagger_{k \uparrow} c^\dagger_{-k \downarrow} c_{k^\prime \uparrow} c_{-k^\prime \downarrow} \\ \nonumber
H_{AF} = \frac{1}{2V_l} \sum_{k k^\prime} U_{k k^\prime} c^\dagger_{k+Q \uparrow} c_{k \downarrow} c^\dagger_{k^\prime \uparrow} c_{k^\prime-Q \downarrow}
\end{eqnarray}

\noindent where $U_{k k^\prime} = U \delta _{kk^\prime} $ is the on-site repulsion with $U = U_{ff}$ for the atomic case and $U = U_e$ for the electronic case. The boson induced effective interaction between fermions $V_{k k^\prime} = V(\vec{k}-\vec{k}^\prime)$ with $V(k) = V_{ph}(k)$ for the atomic case and $V(k) = V_{mag}(k)$ for the electronic case. Here $V_l$ is the volume of the system and $\epsilon_k = -2 t (\cos k_x d + \cos k_y d)-\mu$ with $t = t_f$ and $t = t_e$ for atomic and electronic systems, respectively. While the term $H_0$ in the Hamiltonian represents the kinetic energy of the fermions, the terms $H_{SC}$ and $H_{AF}$ represent the interactions and they are responsible for superconductivity and magnetism, respectively. The Hamiltonian is highly interacting and unable to be solved in the thermodynamic limit even on a high power computer. There are different approximations to tackle this interacting  many-body Hamiltonian by converting it into an effectively non-interacting one. One of the simple and popular approximations is the mean field approximation where the terms with four fermion operators are decoupled into products of quadratic terms. The same results can be obtained by applying so called Hubbard Stratonovich transformation to the Hamiltonian in functional integral method and evaluating the free energy at saddle point level. Here we use the mean field theory where an arbitrary operator $\hat{A}$ is written in the form $\hat{A} = \langle \hat{A} \rangle + \delta \hat{A}$. Here $\delta \hat{A}$ represents the fluctuation around the mean value, $\langle \hat{A} \rangle$. Then a product of two operators $\hat{A}$ and $\hat{B}$ can be written as $\hat{A}\hat{B} \simeq \langle \hat{A} \rangle \hat{B} + \hat{A} \langle \hat{B} \rangle - \langle \hat{A} \rangle \langle \hat{B} \rangle$, where the approximately equal sign comes from neglecting the second order fluctuation term, $\delta \hat{A} \delta \hat{B}$. The exclusion of the second order fluctuation term makes this theory valid only for systems or regimes where quantum fluctuations are unimportant.

In order to convert our Hamiltonian into an effectively non-interacting one, we decouple quartic fermion terms into quadratic terms using the mean field theory. Introducing the two expectation values for fermion bilinear operators as $\Delta_{k^\prime} = \sum_k V_{k k^\prime} \langle c^\dagger_{k \uparrow} c^\dagger_{-k \downarrow} \rangle$ and $M_Q = -U \langle c^\dagger_{k+Q \uparrow} c_{k \downarrow}\rangle$, we write the $H_{SC}$ and $H_{AF}$ in the form,

\begin{eqnarray}
H_{SC} = \frac{1}{2} \sum_{k} \biggr(\Delta_k c_{k \uparrow} c_{-k \downarrow} + \Delta_k^\ast c^\dagger_{k \uparrow} c^\dagger_{-k \downarrow}\biggr) -\frac{1}{2V_l} \sum_{k k^\prime} \frac{\Delta_k^\ast \Delta_{k^\prime}}{V_{k k^\prime}} \\ \nonumber
H_{AF} = -\frac{1}{2} \sum_{k} \biggr(M_Q c^\dagger_{k \uparrow} c_{k-Q \downarrow} + M_Q^\ast c^\dagger_{k+Q \uparrow} c_{k \downarrow}\biggr) +  \frac{M_Q M_Q^\ast}{U}.
\end{eqnarray}

\noindent The non-zero values of the expectation values or the order parameters $\Delta_k$ and $M_Q$ represent superconducting order and magnetic order at wave vector $\vec{Q}$. By introducing a four component vector $\psi^\dagger_k = (c^\dagger_{k \uparrow}, c_{-k\downarrow}, c^\dagger_{k+Q \downarrow}, c_{-k-Q \uparrow})$, the mean field Hamiltonian can be written in the bilinear form,

\begin{eqnarray}
H_{MF} = \psi^\dagger_k D \psi_k  -\frac{1}{2V_l} \sum_{k k^\prime} \frac{\Delta_k^\ast \Delta_{k^\prime}}{V_{k k^\prime}} + \frac{M_Q M_Q^\ast}{U},
\end{eqnarray}

\noindent where $D$ is a $4 \times 4$ matrix given by,

\begin{eqnarray}
D = \left(
      \begin{array}{cccc}
        \epsilon_a + \epsilon_s & 0 & M_Q & \Delta_k \\
        0 & \epsilon_a + \epsilon_s & \Delta_k & M_Q \\
        M_Q & \Delta_k & -\epsilon_a + \epsilon_s & 0 \\
        \Delta_k & M_Q & 0 & -\epsilon_a + \epsilon_s \\
      \end{array}
    \right).
\end{eqnarray}

\noindent Here we defined two parameters $\epsilon_a = (\epsilon_k - \epsilon_{k+Q})/2$ and $\epsilon_s = (\epsilon_k + \epsilon_{k+Q})/2$. By diagonalizing the Matrix $D$, we find the four eigenvalues $E_{l}(k) = \epsilon_s \pm \sqrt{\epsilon_a^2 + (\Delta_k \pm M_Q)^2}$. As formulated above, we have treated that the magnetic order arises due to the on-site repulsive interaction $U$. This is driven by a Fermi surface instability which is predominantly due to the existence of nested Fermi surface. The Fermi surface is nested when its opposite edges are related to one another by a fixed nesting vector $\vec{Q}$ in momentum space. This nesting condition is given by the particle-hole symmetry, $\epsilon_{k+Q} = -\epsilon_k$. However, if the nesting is not perfect then the partially destroyed Fermi surface can allow for the possibility of superconducting instability due to the effective induced interaction. In addition to the nesting condition, the time-reversal symmetry gives $\epsilon_{-k} = \epsilon_k$. In our model, the magnetic instability is anti-ferromagnetic in nature so we use $\vec{Q} = (\pm \pi/d, \pm \pi/d)$. With this nesting condition, the eigenvalues of the Hamiltonian is $E_{\pm}(k) = \pm \sqrt{\epsilon_k^2 + (\Delta_k \pm M_Q)^2}$.

To derive the thermodynamic grand potential, we start with the grand canonical partition function,

\begin{eqnarray}
Z_G = Tr[e^{-\beta H_{MF}}] = \sum_\gamma \langle \gamma|e^{-\beta H_{MF}}|\gamma \rangle,
\end{eqnarray}

\noindent where \emph{Tr} is the trace and the sum goes through the quasiparticle basis $\gamma$. In the quasiparticle basis where quasiparticle occupation numbers $n_{kl} = \langle \gamma^\dagger_{kl}\gamma_{kl} \rangle$ are good quantum numbers, our mean field Hamiltonian has the form,

\begin{eqnarray}
H_{MF} = \sum_k \sum_{l = 1,2,3,4}E_l(k) \gamma^\dagger_{k l} \gamma_{k l} -\frac{1}{2V_l} \sum_{k k^\prime} \frac{\Delta_k^\ast \Delta_{k^\prime}}{V_{k k^\prime}} + \frac{M_Q M_Q^\ast}{U}.
\end{eqnarray}

\noindent where the four eigenvalues $E_l(k)$ are defined as $E_{1,3}(k) = \epsilon_s \pm \sqrt{\epsilon_a^2 + (\Delta_k + M)^2}$ and $E_{2,4}(k) = \epsilon_s \pm \sqrt{\epsilon_a^2 + (\Delta_k - M)^2}$ with upper sign is for the first index and lower sign is for the second index in $E_{l,m}(k)$. The grand partition function then becomes,

\begin{eqnarray}
Z_G = e^{-\beta C} \prod_k \prod_l  \sum_{n_{kl}}e^{-\beta E_{l}(k)n_{kl}}.
\end{eqnarray}

\noindent Using the fact that quasiparticle fermion occupation numbers $n_{kl}$ for a given $k$ are zero and one, and after dropping the unimportant constant term, the thermodynamic grand potential $\Omega = -1/\beta \ln Z_G$ is given by,

\begin{eqnarray}
\Omega = -\frac{1}{\beta} \sum_k \sum_l \ln\biggr[ \cosh (\beta E_l(k)/2)\biggr] - \frac{1}{2V_l} \sum_{k k^\prime} \frac{\Delta_k^\ast \Delta_{k^\prime}}{V_{k k^\prime}} + \frac{M_Q M_Q^\ast}{2 U}.
\end{eqnarray}

\section{V. The Landau Energy Functional for superconductivity and Antiferromagnetism}

The mean-field thermodynamic potential derived in previous section can be used to construct the Landau energy functional for superconducting and anti-ferromagnetic order parameters, $\Delta_k$ and $M_Q$. In order to include the different symmetries of the superconducting order parameter, we take $\Delta_k = \Delta_\eta \eta_k$, where $\eta_k =1$ for the spin-singlet on-site s-wave pairing, $\eta_k = 2[\cos k_xd + \cos k_yd]$ for the spin-singlet extended off-site s-wave pairing, and $\eta_k = 2[\cos k_xd - \cos k_yd]$ for the extended off-site d-wave pairing. For the anti-ferromagnetic order parameter where $\vec{Q} d = \{\pi, \pi\}$, we assume $M_Q = M$. Since the thermodynamic potential is analytic at both $\Delta_k = 0$ and $M_Q = 0$ at finite temperatures, we can expand it to the quartic order to get our Landau energy functional in the form,

\begin{eqnarray}
F_{LG} = \frac{1}{2} \alpha_s \Delta_\eta^2 + \frac{1}{4} \beta_s \Delta_\eta^4 + \frac{1}{2} \alpha_m M^2 + \frac{1}{4} \beta_m M^4 + \frac{1}{2} \gamma \Delta_\eta^2M^2.
\end{eqnarray}

\noindent Here we have neglected the higher order terms and for the stability of the Landau energy functional, both $\beta_s$ and $\beta_m$ must be positive. In addition, the parameter $\gamma$ is restricted to the region $\gamma > -\sqrt{\beta_s \beta_m}$ for the energy to be bounded from below. A similar Landau expansion is investigated in Refs.~\cite{lee1, lee2, lee3, lee4}. We restrict ourselves to the parameter regime where these conditions are satisfied. Then the Landau coefficients $\alpha_s = \frac{\partial^2 \Omega}{\partial \Delta_\eta^2}|_{\Delta_\eta= 0, M = 0}$, $\alpha_m = \frac{\partial^2 \Omega}{\partial M^2}|_{\Delta_\eta= 0, M = 0}$, $\beta_s = \frac{1}{6}\frac{\partial^4 \Omega}{\partial \Delta_\eta^4}|_{\Delta_\eta= 0, M = 0}$, $\beta_m = \frac{1}{6}\frac{\partial^4 \Omega}{\partial M^4}|_{\Delta_\eta= 0, M = 0}$, and $\gamma = \frac{1}{2}\frac{\partial^4 \Omega}{\partial \Delta_\eta^2 M^2}|_{\Delta_\eta= 0, M = 0}$ are derived from the mean-field thermodynamic potential $\Omega$. The explicit expressions for these coefficients are given in the appendix. The stable thermodynamic phases are determined by the values of the order parameters. The non-zero order parameter suggests the ordered phase, hence the model can predict four different thermodynamic phases, normal ($\Delta_\eta =0, M = 0$), superconducting ($\Delta_\eta \neq 0, M = 0$), anti-ferromagnetic ($\Delta_\eta = 0, M \neq 0$), and the co-existing phase of anti-ferromagnetism and superconductivity or the mixed phase ($\Delta_\eta \neq 0, M \neq 0$). The theory is valid close to the critical temperatures where the order parameters are small. The phase transition is determined by the sign of $\alpha_{s, m}$. When $\alpha_{s, m}$ changes signs from positive to negative for a given set of system parameters, the system enters from the normal state to an ordered state. The detail investigation of the phase diagram within this Landau approach is given in sections VII below.

\section{VI. The gap equations and linearization}

The gap equations for the two order parameters, $\Delta_k$ and $M_Q$ are obtained from minimizing the mean-field thermodynamic potential $\Omega$. The minimization, $\partial \Omega/\partial \Delta_k = 0$ and $\partial \Omega/\partial M = 0$ leads to

\begin{eqnarray}
\Delta_k = - \sum_{q} \biggr\{\frac{V_{kq} (\Delta_q + M)}{2E_{+}(q)} \biggr(\tanh (\beta E_{1}(q)/2) - \tanh (\beta E_{3}(q)/2) \biggr) \\ \nonumber  -  \frac{V_{kq} (\Delta_q - M)}{2E_{-}(q)} \biggr(\tanh (\beta E_{2}(q)/2) - \tanh (\beta E_{4}(q)/2) \biggr)\biggr \}
\end{eqnarray}

\noindent and

\begin{eqnarray}
M =   \frac{U}{2}\sum_{q} \biggr\{\frac{(\Delta_q + M)}{E_{+}(q)} \biggr(\tanh (\beta E_{1}(q)/2) - \tanh (\beta E_{3}(q)/2) \biggr) \\ \nonumber  +  \frac{ (\Delta_q - M)}{E_{-}(q)} \biggr(\tanh (\beta E_{4}(q)/2) - \tanh (\beta E_{2}(q)/2) \biggr)\biggr \}.
\end{eqnarray}

\noindent In principle, these two non-linear equations must be solved self consistently for the order parameters. These self-consistency conditions demand considerable numerical efforts. However, the thermal phase transitions can be determined by linearized gap equations which are valid close to the critical temperatures where the order parameters are small. By expanding the gap equations around $\Delta_k = 0$ and $M =0$, and keeping only the linear order, we have two linearized gap equations,

\begin{eqnarray}
\Delta_k = -  \sum_{q} V_{kq} \Delta_q S_q \enspace \hbox{and} \enspace M =  UM \sum_{q} S_q,
\end{eqnarray}

\noindent where we defined, $S_q = [\tanh (\beta \epsilon_k/2)- \tanh (\beta \epsilon_{k+Q}/2)]/\epsilon_a$. In order to include both s-wave and d-wave superconducting symmetries, we expand the superconducting order parameter using the ansatz $\Delta_k = \Delta_0 + \Delta_s \gamma_k + \Delta_d \theta_k$. Here we choose the base functions $\gamma_k = 2 [\cos k_x d + \cos k_y d]$ and $\theta_k = 2 [\cos k_x d -\cos k_y d]$ so that $\Delta_0$, $\Delta_s$, and $\Delta_d$ represent on-site s-wave, off-site s-wave, and off-site d-wave order parameters. Inserting this ansatz into the linearized gap equation, we construct three equations,

\begin{eqnarray}
(A - 1) \Delta_0 + B \Delta_s + C \Delta_d = 0 \nonumber \\
D \Delta_0 + (E - 1) \Delta_s + F \Delta_d = 0 \nonumber \\
G \Delta_0 + H \Delta_s + (I - 1) \Delta_d = 0.
\end{eqnarray}

The equations are derived from the linearized superconducting gap equation. All the coefficients from A through I are listed in the appendix. The first equation is derived by summing the linearized gap equation over momentum $k$. The second equation is derived by, first multiplying linearized equation by $\gamma_k$ and then summing over the momentum. The third equation is obtained, first by multiplying $\theta_k$ and then summing over the momentum. We solve these three equations simultaneously for the superconducting order parameter and then use AFM gap equation for anti-ferromagnetic order parameter as we discuss in section VIII below.

\section{VII. The generic phase diagram from the Landau Energy Functional Approach}

\begin{figure}
\includegraphics[width=\columnwidth]{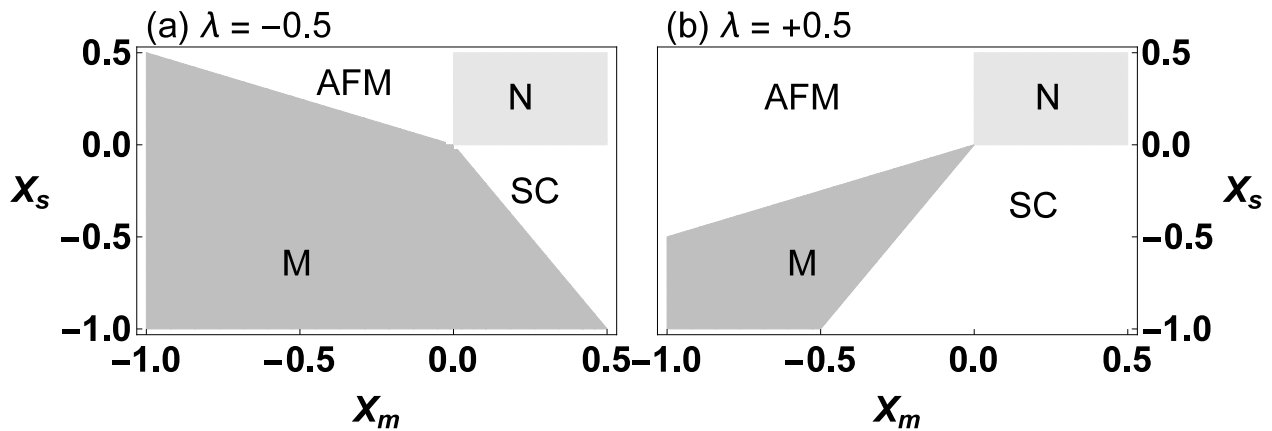}
\caption{The generic phase diagram originated from the Landau energy functional in the text. The panel (a) is for parameter $\lambda = -0.5$ and panel (b) is for $\lambda = + 0.5$. In $X_s$ and $X_m$ parameter space, thermodynamically stable phases are denoted by AFM: anti-ferromagnetic, SC-superconducting, N: normal (neither AFM nor SC), and M: mixed (simultaneous existence of AFM and SC). See text for the details.}\label{PDX}
\end{figure}

As we discussed in section V above, the Landau energy functional predicts four different phases depending on the parameters $\alpha_{s, m}$, $\beta_{s, m}$, and $\gamma$. We find that the generic phase diagram can be constructed within a three-parameter space given by $\lambda = \gamma/\sqrt{\beta_s \beta_m}$, $X_s = \alpha_s/\sqrt{\beta_s}$, and $X_m = \alpha_m/\sqrt{\beta_m}$. By analyzing the energy functional and the order parameters $\Delta$ and $M$ as usual (\emph{ie}, by minimizing the energy functional and then solving the minimized equations simultaneously for real order parameters), we find that only the disordered normal phase is stable for $X_s >0$, $X_m >0$, and $\lambda > -1$. For $\lambda < -1$, the free energy is unbounded from below for all values of $X_s$ and $X_m$. As expected, $X_s <0$ is necessary for the superconducting phase and $X_m <0$ requires for the anti-ferromagnetic phase. However, depending on the value of $\lambda$, the co-existence of superconductivity and anti-ferromagnetism or the mixed phase can be thermodynamically stable if one of the parameters $X_s$ and $X_m$ or both are negative. Table \ref{T1} summarizes the energy, the order parameters, and the parameter space for the phase diagram of our Landau energy functional. In order to determine the thermodynamically stable phase, we searched not only the lowest energy, but also the non-imaginary order parameters. The generic phase diagram for two representative values of $\lambda$ is shown in FIG. \ref{PDX}.

\begin{table*}[t!]
\begin{center}
\begin{tabular*}{0.64\linewidth}{|l|l|}
  \hline
\textbf{PHASE} & \textbf{PARAMETER RANGE} \\
  \hline
  Normal (N) & \\
  \hline
  $F_N = 0$ & $X_s >0$, $X_m >0$ \\
  $\Delta_\eta^2 = M^2 =0$ &  \\
  \hline
 Ant-ferromagnetic (AFM) &  \\
 \hline
 $F_{AFM} = -\frac{1}{4} X_m^2$ & (a).  $X_m < 0$, $X_s > 0$, $\lambda > \frac{X_s}{X_m}$, $\lambda > max\biggr(-1, \frac{X_m}{X_s}\biggr)$ \\
 $\Delta_\eta^2 = 0$ &  \\
 $M^2 = -\frac{X_m}{\sqrt{\beta_m}}$ & (b). $X_m < 0$, $X_s < 0$, $\lambda > \frac{X_m}{X_s}$ \\
 \hline
 Superconducting (SC) &  \\
 \hline
 $F_{SC} = -\frac{1}{4} X_s^2$ & (a). $X_m > 0$, $X_s < 0$, $\lambda > \frac{X_m}{X_s}$, $\lambda > max\biggr(-1, \frac{X_s}{X_m}\biggr)$ \\
 $\Delta_\eta^2 = -\frac{X_s}{\sqrt{\beta_s}}$ &  \\
 $M^2 = 0 $ & (b). $X_m < 0$, $X_s < 0$, $\lambda > \frac{X_s}{X_m}$ \\
 \hline
 Mixed (M) &   \\
 \hline
 $F_{M} = \frac{X_s^2 + X_M^2 - 2\lambda X_s X_m}{4(\lambda^2-1)}$ &  (a). $X_m < 0$, $X_s > 0$, $max\biggr(-1, \frac{X_m}{X_s}\biggr) < \lambda < \frac{X_s}{X_m}$ \\
 $\Delta_\eta^2 = \frac{1}{\sqrt{\beta_s}} \frac{X_s - \lambda X_m}{\lambda^2-1}$ & (b). $X_m > 0$, $X_s < 0$, $max\biggr(-1, \frac{X_s}{X_m}\biggr) < \lambda < \frac{X_m}{X_s}$ \\
 $M^2 = \frac{1}{\sqrt{\beta_m}} \frac{X_m - \lambda X_s}{\lambda^2-1}$  &  (c). $X_m < 0$, $X_s < 0$, $-1 < \lambda < min\biggr(\frac{X_s}{X_m}, \frac{X_m}{X_s} \biggr)$ \\
  \hline
\end{tabular*}
\caption{The phase boundary, energy, and the order parameters of thermodynamically stable phases from the generic Landau energy functional.}\label{T1}
\end{center}
\end{table*}

For a two dimensional square lattice, the nesting is known to occur at the wave vector $\vec{Q}d = (\pi, \pi)$ for half filled fermions~\cite{nes1, nes2}. This leads to a divergence of $\chi(\vec{Q}, T\rightarrow 0)$ at half filling. Further, for a two dimensional square lattice, one finds a spin fluctuation peak at the AFM wave vector $\vec{Q}d = (\pi, \pi)$, even away from zero temperature and half filling limits. Indeed, the inelastic neutron scattering measurements have shown a spin resonance peak in cuprates at the AFM wave vector~\cite{ns1,ns2}. As the momentum dependence on the pairing interaction $V_{k k^\prime} = V(\vec{k}-\vec{k}^\prime)$ is mainly determined by the momentum dependence of the susceptibility $\chi_{\{B, 0\}}(\vec{q})$, we evaluate the pairing interaction at the AFM wave vector $\vec{Q}$ for all our calculations.

\begin{figure*}
\includegraphics[width=\columnwidth]{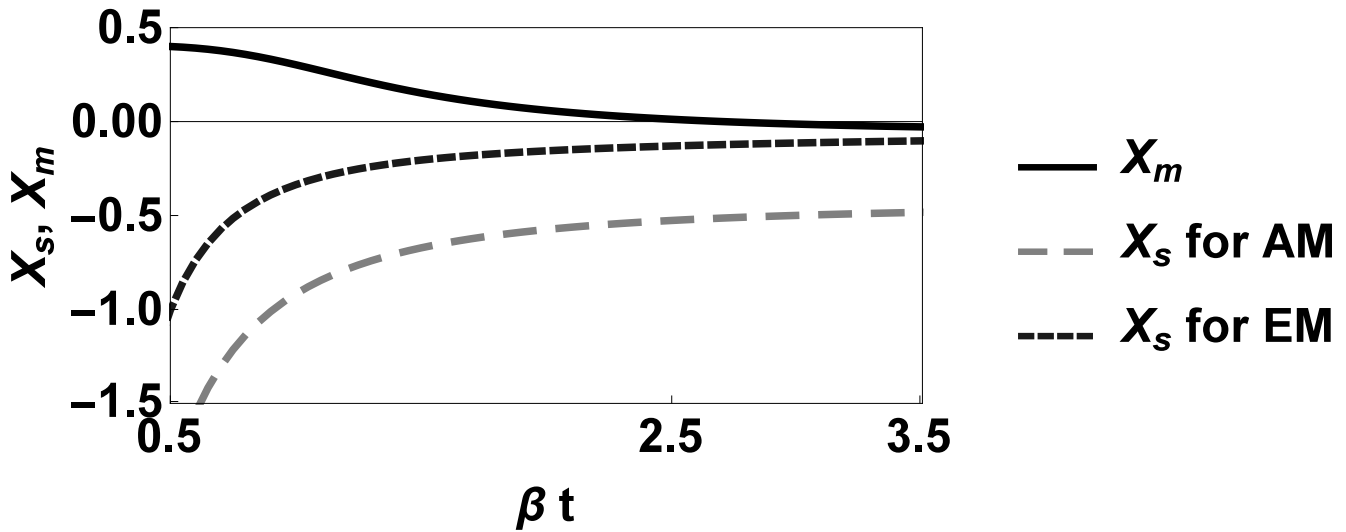}
\caption{The value of $X_m$ (black) and $X_s$ (gray) for both atomic mixture (AM) and electronic model (EM) as a function of dimensionless inverse temperature $ \beta t$. We fixed the parameters as $ \mu =0$ and $U = 2 t$ for both cases. The value of $X_m$ is same for both systems, the value of $X_s$ is different due to the different forms of the attractive interaction. For the atomic mixture, we set $t_b = 2 t$, $n_b =2$, $U_{bb} = t$, and $U_{bf} = 3 t$.}\label{XmXs}
\end{figure*}

Notice that $\beta_s$ and $\gamma$ given in the appendix depend only on the symmetry of the superconducting order parameter $\eta_k$ but not on the interaction. As a result, $\lambda$ and $X_m$ are the same for both Bose-Fermi and electronic systems. For all values of $\mu$ and $\beta$, we find that $\lambda < -1$ for both d-wave pairing and on-site s-wave pairing. This alone does not rule out the possibility of d-wave and on-site s-wave superconductivity, but one has to consider the higher order terms in the Landau energy functional that we neglected in our calculations. By evaluating the integral numerically, we calculate $X_s$ and $X_m$ as a function of chemical potential ($\mu$), inverse temperature ($\beta$), and interaction ($U$). As a demonstration we show both $X_s$ and $X_m$ as a function of chemical potential $\mu$ for a chosen set of parameters relevant to both electronic model and Bose-Fermi mixture in FIG~\ref{XmXs}. In general, the chemical potential controls the doping level or the filling factors. In cold atom experiments, atoms are trapped using a combined harmonic oscillator trapping potential and optical lattice potential. As a result of the harmonic oscillator trapping potential, the atomic density is not homogeneous in the lattice. The number of Fermi atoms decreases as one goes from the center to the edge of the trap. This results in the chemical potential monotonically decreasing from the center to the edge of the trap. Therefore, $X_s$ and $X_m$ values shown in FIG. \ref{XmXs} show the variation of those values in real space. We find that both $X_s$ and $X_m$ can have both negative and positive values depending on the system parameters showing the possibility of having all the phases discussed in FIG. \ref{PDX}. For the electronic model, the chemical potential controls the external career concentration which is generally induced by the doping of parent high temperature materials by external atoms.

\section{VIII. Phase transition from linearized theory}

The linearized version of the gap equations derived in section VI (Eq. 23) can be written in compact form as a matrix equation, $\bar{M} \vec{\Delta} = 0$, where $\bar{M}$ is a $3 \times 3$ matrix and $\vec{\Delta} = \{\Delta_0, \Delta_s, \Delta_d\}^{t_s}$ is a 3-component column vector with $t_s$ being the transpose. Then the critical temperature is determined by the condition $\det \bar{M} = 0$ (\emph{i.e}, setting the determinant of matrix $\bar{M}$ to be zero) . All nine matrix elements are related to the parameters $A-I$ listed in appendix. However, due to the nature of function $\theta_k$ and symmetry of the integral ($k_x \Leftrightarrow k_y$) we have only five non-zero matrix elements to be calculated (we find $C = F= G = H = 0$ and $D = -B/4$). Then the condition $\det \bar{M} = 0$ leads to two equations,

\begin{eqnarray}
I -1 =0 \enspace \hbox{and} \enspace AE + B^2/4-A - E +1 =0.
\end{eqnarray}

\noindent The critical parameters of the superconducting phase are determined by the solution of these two equations. While the first equation determines critical parameters for d-wave pairing, the second one determines that of s-wave pairing. By solving these two equations and the linearized magnetic gap equations numerically for given values of on-site interaction $U$, we find the critical temperatures of both s-wave and d-wave pairing, and anti-ferromagnetic transition as a function of chemical potential for both atomic mixture and electronic model.

\begin{figure*}
\includegraphics[width=\columnwidth]{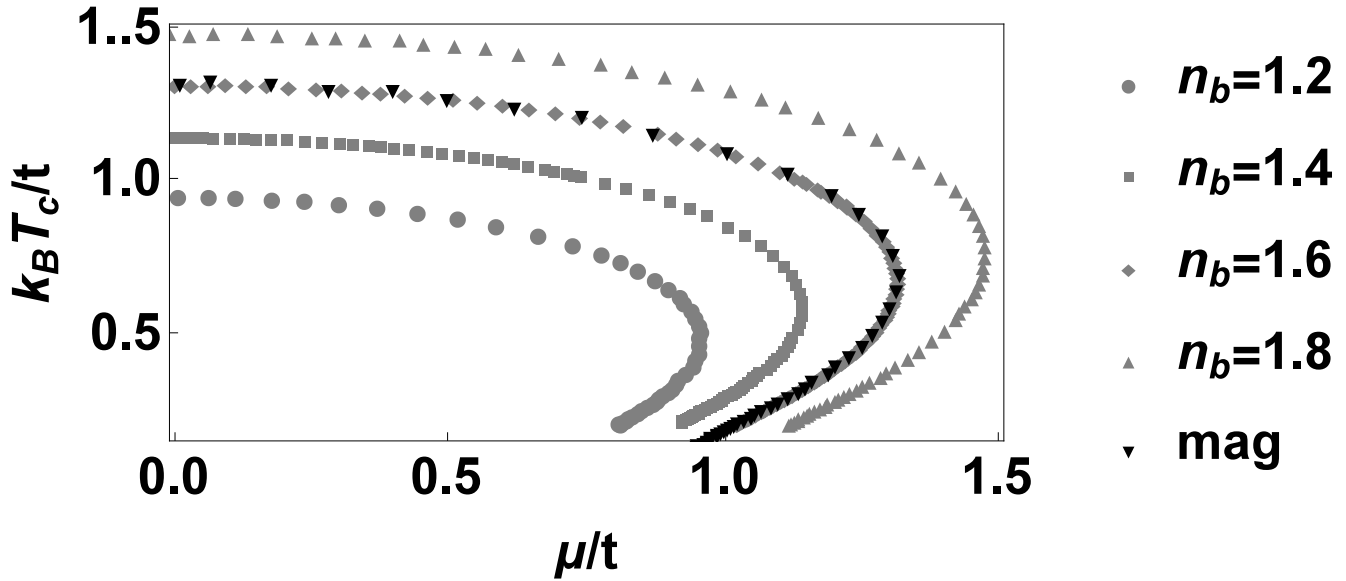}
\caption{Critical temperatures $k_BT_c/t$ as function of chemical potential $\mu$ for the atomic mixture. These are calculated from the solutions of linearized gap equations as discussed in the text. The s-wave pairing and anti-ferromagnetic magnetic transitions are shown as a function of chemical potential $\mu$. We set $U = 1.5 t$, $t_b = 2 t$, $U_{bb} = t$, and $U_{bf} = 3t$, but varies the boson density $n_b$ from 1.2 to 1.8 (from bottom to top, gray symbols). For the atomic mixture, we do not find any solutions for d-wave pairing transitions so that the d-wave pairing is absent within the parameters we searched.}\label{PDAM}
\end{figure*}

As we discussed in section II, the attractive interaction between fermions induced by the Bose condensed atoms depends on the Bogoliubov spectrum of the condensed bosons, boson density, and on-site Bose-Fermi interaction. The Bogoliubov spectrum is a function of boson tunneling amplitude and the on-site Bose-Bose interaction. As a result, the superconducting critical temperature depends on all these parameters. We seek solutions for our linearized equations by searching an experimentally relevant large parameter region, however we do not find any indication of d-wave pairing for the atomic mixture. This is not surprising as it is well known that fermions favor s-wave pairing in the limit of a short healing length. The s-wave pairing critical temperature and the magnetic transition temperature are plotted in FIG. \ref{PDAM} for a set of representative parameters. As can be seen from the FIG. \ref{PDAM}, the solutions of linearized equations have two-branch structure. However only the upper branch can be considered as the phase boundary as the lower branch is already reside below the critical temperature. Note that the s-wave critical temperature can be easily controlled to be above or below the magnetic phase transition temperature by varying the boson density. For the chosen parameters in the FIG. \ref{PDAM}, the s-wave superconducting phase transition and anti-ferromagnetic phase transition are simultaneous at the boson density $n_b = 1.8$. As one expects, the atomic system is in the normal phase at larger chemical potentials and larger temperatures, however it can be in the superconducting phase, magnetic phase, or co-existing phase at lower chemical potentials depending on the temperature and the other system parameters such as boson density and boson tunneling energy.

 \begin{figure*}
\includegraphics[width=\columnwidth]{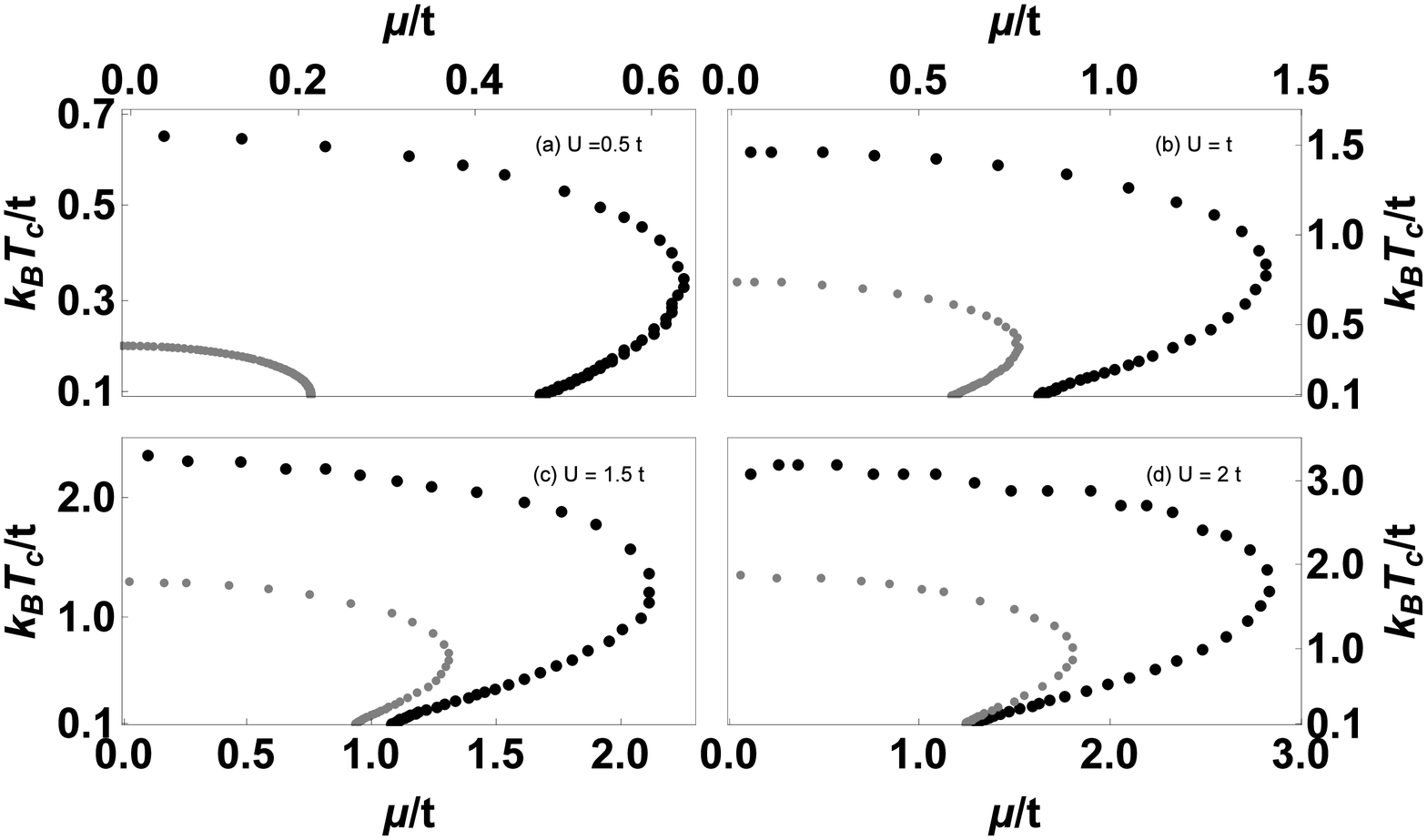}
\caption{Critical temperatures $k_BT_c/t$ as function of chemical potential $\mu$ for the electronic model. These are calculated from the solutions of linearized gap equations as discussed in the text. Each panel shows four different on-site interactions $U$. While black dots represent the d-wave pairing transition temperatures, gray dots represent the anti-ferromagnetic transition temperature. For our electronic model, we do not find any solutions for s-wave pairing transitions so that the s-wave pairing is absent within the parameters we searched.}\label{PDEM}
\end{figure*}

Unlike the Bose-Fermi mixture, the pairing interaction between electrons in our electronic model depends on the on-site interaction. In addition, the magnetic transition also depends on the on-site electron-electron interaction. Therefore, we search the solutions for our linearized equations for three different regimes, namely weak coupling ($U < t$), intermediate coupling ($U \sim t$), and strong coupling ($ U > t$). We restrict the search to a reasonable experimentally relevant parameters, $0 \leq \mu/t \leq 2$ and $0 \leq \beta t \leq 10$. The qualitative behavior of the solutions are same in all three regimes. The solutions have two-branch structure similar to those of the Bose-Fermi case. In all three regimes of our electronic model, we find s-wave pairing is absent within the experimentally relevant parameter range we searched. However, the d-wave pairing between fermions can take place as we show d-wave critical temperatures in FIG.~\ref{PDEM} in all three regimes. For the entire parameter regime searched, the anti-ferromagnetic transition is the lowest. As one increases the on-site interaction, all transition temperatures increase as expected, however the anti-ferromagnetic transition temperature increases at a faster rate than that of the d-wave pairing temperature. As a result, anti-ferromagnetic phase transition may take place first as one decreases the temperature at extremely larger on-site interactions. Similar to the atomic system, the electronic system also in the normal phase at larger chemical potentials and larger temperatures, however at smaller chemical potentials at low enough temperatures, the system is in either d-wave superconducting phase or coexisting phase unless $U \rightarrow \infty$.

\section{IX. Discussions and Summary}

We have considered both an electronic model and a cold atom mixture in a square lattice to study the interplay between superconductivity and anti-ferromagnetism. We assume that the atom mixture is made up of a two-component Fermi gas and a single component Bose gas where the bosons are in Bose-Einstein condensation. In addition to the tunneling and on-site interaction of fermions, an elementary excitation mediated attractive interaction at the nesting wave-vector is also considered for both systems. While spin fluctuations are taken as the attractive mediators for the electronic system, phonon excitations of condensed bosons are taken as the attractive mediators for the atom mixture on optical lattices. Then focusing on an effective model within a mean field theory, we have explored superconductivity and anti-ferromagnetism in the systems. We treated both s-wave pairing and d-wave pairing, and anti-ferromagnetism at equal footing to study the phase transition by solving linearized gap equations and the Landau energy approach.

First, we studied a general two order parameter Landau energy functional and constructed the generic phase diagram within a three-parameter space. Then calculating relevant parameters for both atomic and electronic models, we find that both anti-ferromagnetic and off-site s-wave superconducting phases simultaneously co-exist in certain parameter regions. Within this Landau approach however, we do not find d-wave pairing or on-site s-wave pairing of fermions.

Second, we studied the phase transitions of both atomic and electronic models by solving the linearized gap equations for both superconductivity and anti-ferromagnetism. For the Bose-Fermi mixture, we do not find the d-wave pairing transition, however we find anti-ferromagnetic and s-wave superconducting phase transitions as one tunes the system parameters. For the electronic model, we find anti-ferromagnetic and d-wave superconducting phase transitions but not s-wave pairing transitions.

Although we have focused on superconducting compounds with spin fluctuation mediated electron-electron attraction, our qualitative results are applicable to other superconducting compounds, such as iron chalcogenides~\cite{icsc}, organic superconductors~\cite{osc} and heavy-fermions~\cite{hfsc}. However, the pairing of electrons in those compounds can originate from a different mechanism and the external mechanical pressure may play the role of doping. Even though iron pnictide and iron chalcogenide superconducting compounds show similar structures, angle-resolved photoemission spectroscopy on iron chalcogenide displays \emph{only} electron-like pockets on the Fermi surface~\cite{arp1, arp2}. Therefore, Fermi surface nesting condition discussed in the present work may not be applicable for iron chalcogenides. The phase transitions we discussed in the present study qualitatively share a similar experimental phase transitions with cuprates, iron pnictides, organic superconductors, and heavy-fermions~\cite{ept1, ept2}. In electronic matter, these phase transitions have been probed using muon spin relaxation and neutron scattering measurements~\cite{mue,nue}. The cold-atom setup studied in the present paper provides platforms for deeper understanding of the magnetic and superconducting phases found in this electronic matter. Anti-ferromagnetism of fermions in optical lattices have already been probed~\cite{riceE} and detected using Bragg scattering of photons~\cite{bs1, bs2}. The superfluidity of fermion pairs can be detected using photoassociation spectroscopy where weakly bound Cooper pairs are converted into molecules using laser induced transitions~\cite{lit1, lit2}. The symmetry of the superfluid state may be probed using density-density correlation~\cite{oldt7}.

In conclusion, we have studied an experimentally feasible tight-binding effective fermion Hamiltonian relevant for both electronic matter and cold atom setups to investigate the interplay between anti-ferromagnetism and superconductivity. We used two different approximate schemes within a mean-field theory and find the possibility of having both anti-ferromagnetic and superconducting phases as well as the co-existence of these phases with certain parameters.

\section{X. Acknowledgments}
We are very grateful to Andreas Eberlein for critical comments on the first draft of this manuscript.

\section{Appendix}

Here we present the coefficients of the Landau energy functional discussed in section V. As we have discussed in the main text, these are derived from the mean-field thermodynamic potential,

\begin{eqnarray}
\alpha_s = -\sum_{k,q} \biggr\{\frac{\eta_k \eta_q}{V_{kq}} + \sum_{\pm} \frac{2 \eta_k^2}{\epsilon_a} \tanh [\beta (\epsilon_a \pm \epsilon_s)/4]\}  \\
\alpha_m = \frac{1}{U} -\sum_{k} \sum_{\pm} \biggr\{ \frac{2}{\epsilon_a}\tanh [\beta (\epsilon_a \pm \epsilon_s)/4]\biggr\} \nonumber \\
\beta_s = \sum_k \sum_{\pm} \frac{\eta_k^4}{\epsilon_a^3} \{-6 \beta \epsilon_a sech^2[\beta(\epsilon_a \pm \epsilon_s)/4] + 24 \tanh[\beta(\epsilon_a \pm \epsilon_s)/4]\} \nonumber \\
\beta_m = \sum_k \sum_{\pm} \frac{1}{\epsilon_a^3} \{-6 \beta \epsilon_a sech^2[\beta(\epsilon_a \pm \epsilon_s)/4] + 24 \tanh[\beta(\epsilon_a \pm \epsilon_s)/4]\} \nonumber \\
\gamma = \sum_k \sum_{\pm} \frac{6 \eta_k^2}{\epsilon_a^3}\{\beta \epsilon_a sech^2[\beta(\epsilon_a \pm \epsilon_s)/4] - 4 \tanh[\beta(\epsilon_a \pm \epsilon_s)/4]\}, \nonumber
\end{eqnarray}

\noindent where $\pm$ sum needed to be completed with the upper sign for $+$ and the lower sign for $-$. Following are the coefficients of linearized gap equations discussed in section VI. As we have discussed in the main text, some of these coefficients are zero due to nature of the functions $\gamma_k$ and $\theta_k$, and (anti)symmetry of the functions ( $\gamma_k$) $\theta_k$ under $k_x \rightarrow k_y$.

\begin{eqnarray}
A = -\sum_{kq} V_{kq} S_q,  \enspace
B = -\sum_{kq} V_{kq} S_q \gamma_q   \\
C = -\sum_{kq} V_{kq} S_q \theta_q, \enspace
D = -\sum_{kq} V_{kq} S_q \gamma_k/4 \nonumber \\
E = -\sum_{kq} V_{kq} S_q \gamma_k \gamma_q/4, \enspace
F = -\sum_{kq} V_{kq} S_q \gamma_k \theta_q/4 \nonumber \\
G = -\sum_{kq} V_{kq} S_q \theta_k/4, \enspace
H = -\sum_{kq} V_{kq} S_q \gamma_q \theta_k/4 \nonumber \\
I = -\sum_{kq} V_{kq} S_q \theta_q \theta_k/4 \nonumber
\end{eqnarray}

\end{document}